\documentclass[letterpaper]{article} 
\usepackage[]{aaai25}  
\usepackage{xspace}
\usepackage{multirow}
\usepackage{amsthm}
\usepackage{amssymb}
\usepackage{booktabs}
\usepackage{times}  
\usepackage{helvet}  
\usepackage{courier}  
\usepackage[hyphens]{url}  
\usepackage{graphicx} 
\usepackage{natbib}  
\usepackage{caption} 
\usepackage{xcolor}
\definecolor{ballblue}{rgb}{0.13, 0.67, 0.8}
\definecolor{forestgreen}{rgb}{0.13, 0.55, 0.13}
\definecolor{amber}{rgb}{1.0, 0.75, 0.0}
\definecolor{salmonpink}{rgb}{1.0, 0.57, 0.64}
\definecolor{blue-violet}{rgb}{0.54, 0.17, 0.89}
\definecolor{katscolor}{RGB}{0, 167, 159}

\theoremstyle{plain}
\newtheorem{theorem}{Theorem}

\theoremstyle{definition}
\newtheorem{definition}[theorem]{Definition}

\theoremstyle{remark}

\usepackage{algorithm}
\usepackage{algorithmic}

\newcommand{\taxonomy}{taxonomy\xspace}

\newcommand{\refTax}[2]{V#1.#2/I#1.#2\xspace}
\newcommand{\refVuln}[2]{V#1.#2\xspace}
\newcommand{\refInc}[2]{I#1.#2\xspace}

\newcommand*\rotvertical{\rotatebox{90}}

\usepackage{newfloat}
\usepackage{listings}
\DeclareCaptionStyle{ruled}{labelfont=normalfont,labelsep=colon,strut=off} 
\floatstyle{ruled}
\newfloat{listing}{tb}{lst}{}
\floatname{listing}{Listing}
\title{Practice-Informed, Practice-Ready: An AI security incident taxonomy}

\author{
    Lukas Bieringer\textsuperscript{\rm 1},
    Sean McGregor\textsuperscript{\rm 2},
    Nicole Nichols\textsuperscript{\rm 3},
    Kevin Paeth\textsuperscript{\rm 4},
    Andrew Paverd\textsuperscript{\rm 5},\\
    Jonathan Petit\textsuperscript{\rm 6},    
    Jochen Stängler\textsuperscript{\rm 7},
    Andreas Wespi\textsuperscript{\rm 8},
    Alexandre Alahi\textsuperscript{\rm 9},
    Kathrin Grosse\textsuperscript{\rm 10}
}
\affiliations{
    \textsuperscript{\rm 1}QuantPi, Saarbr\"ucken, Germany, 
     \textsuperscript{\rm 2}Responsible AI Collaborative, US 
    \textsuperscript{\rm 3}Palo Alto Networks, US  \\
    \textsuperscript{\rm 4}UL Research Institutes, US 
    \textsuperscript{\rm 5}Microsoft Security Response Center, UK,
    \textsuperscript{\rm 6}Qualcomm, US, 
    \textsuperscript{\rm 7}German Federal Office \\ for Information Security, BSI, Germany,
    \textsuperscript{\rm 8}IBM Research Zurich, Switzerland, 
    \textsuperscript{\rm 9}EPFL VITA Lab, Switzerland \\
    \textsuperscript{\rm 10}Independent Researcher \hspace{2em}
    kathrgrosse@gmail.com
}

\begin{document}

\maketitle

\begin{abstract}
With the increasing prevalence of AI systems, several real-world AI security incidents have been reported.
However, despite forthcoming legal mandates, the reporting and collection of these incidents still lacks practical standards and proposals.
We bridge this gap by establishing a rigorous foundation based on discussions with a diverse group of AI practitioners spanning industrial, non-profit, research, and governmental sectors. Our proposed taxonomy provides concrete guidance to identify affected parties, recommend relevant security measures, and gain an actionable overview of the evolving AI security landscape. Our tests show that different coders consistently identify similar topics, but that automating incident tagging via an LLM like ChatGPT is of limited use. Notably, our framework has already served as the scientific basis for an established industry standard, proving its utility and readiness for widespread adoption.
\end{abstract}

%

\section{Introduction}
AI adoption is on the rise: a third of the adult population in Western countries use large language models~\cite{OECD2026AI}, and industrial usage is common. At the same time, a plethora of attacks or manipulations of AI have been shown~\cite{biggio2018wild, cina_wild_2022,lei2023new,oliynyk2023know,shen2023anything,shumailov2021manipulating,yao2024survey}. 
There have also been reports that these attacks can affect real-world systems~\cite{grosse2023your,mcgregor2021preventing,avid,shen2023anything}. Some notable examples include: malicious AI dependencies\footnote{\url{https://avidml.org/database/}, 2023-V015}, bypassed malware detectors\footnote{\url{https://avidml.org/database/}, 2023-V004}, reverse engineering of existing models~\cite{grosse2023your}, search engine poisoning\footnote{\url{https://incidentdatabase.ai/cite/88}}, and poisoning of large language models~\cite{carlini2024poisoning}.

However, there operational usefulness of reported incidents is limited~\cite{grosse2023your} as there is no guidance for organizations \emph{how} to describe AI security incidents, as existing approaches do not fully meet the unique needs of the AI security domain~\cite{bieringer2024position}.
Yet, reporting and analyzing security incidents related to AI systems is a fundamental prerequisite for adequate risk management. Regulations increasingly require such reporting, for example, Article 55 (1c) of the~\citet{madiega2021artificial} AI Act requires reporting of \emph{``relevant information about serious incidents and possible corrective measures''}. 
More recently, additional specific commitments regarding serious incident reporting have been published in the EU's General-Purpose AI Code of Practice\footnote{\url{https://digital-strategy.ec.europa.eu/en/library/second-draft-general-purpose-ai-code-practice-Dpublished-written-independent-experts}}, where providers of general-purpose AI systems will be obliged to report relevant incident information such as affected models, root causes, or fail-safe measures when serious incidents occur.

We present an operational taxonomy for AI security incidents developed through iterative input and feedback. Among our authors and feedback providers, there are voices from industry, non-profit, governmental and academic organizations. 
We distilled these voices into a practical reporting scheme for AI security incidents.  
Our taxonomy has been adopted as the basis for the ETSI reporting standard for AI common incident expressions.  

Our taxonomy organizes reporting of the \textit{AI security incidents} by a similar split as in cybersecurity, where we additionally suggest a description of the underlying \textit{AI vulnerabilities}. Both are defined using established standards and definitions. Incidents and vulnerabilities have an $n$ to $m$ relationship: one vulnerability may cause several incidents, or several vulnerabilities may lead to one incident. In both cases, we suggest a naming scheme and required meta information. Further information supports affected party notification by keeping track of exact model and usage. In addition, the vulnerability origin in the supply chain has to be tracked to enable efficient notification of affected parties. Lastly, collecting information about the system context allows to determine whether the incident potentially generalizes to other AI systems.
Although our taxonomy includes a maximal set of relevant information, it is clear that some of this information may be considered optional in some cases.

Our aims with this work are 1) to support organizations in documenting and reporting AI security incidents, and 2) to fuel discussion within the broader community around AI security reporting requirements and best practices.

\section{Related Work}
To the best of our knowledge, there is no established procedure for AI security incident reporting. 
There are incident reporting frameworks surrounding AI that consider non-security incidents, for example, autonomous vehicles accidents due to malfunction.\footnote{\url{https://www.sf.gov/report-autonomous-vehicle-incidents}} There are also several databases collecting general public AI incidents~\cite{oecdaim,avid,mcgregor2021preventing}.
More closely related, \citet{cattell2024coordinated} propose a CVE-like framework for adjudicating AI flaws more generally (e.g., incident reporting for undesired behaviors of AI systems) as well as other improvements (such as extended model cards). As flaws are a more general concept not requiring an adversary or vulnerability to be reportable, \citet{cattell2024coordinated} is orthogonal to our work. 
In addition, \citet{fazelnia2024establishing} proposed and MITRE announced a general AI incident sharing initiative\footnote{\url{https://ai-incidents.mitre.org}} concurrently to us. While both proposals bear similarities to ours, we suggest a more detailed treatment of the intended purpose, supply chains, and existing security measures. \citet{fazelnia2024establishing} also introduces an AI bill of materials (AIBOM) summarizing all components of AI models~\cite{bennet2024AIBOM}. Our \taxonomy is extensible with an AIBOM standard. Also, \citet{ezell2025incident} proposed an incident collection, but focusing on non-security AI agent incidents.

\section{Approach and Verification}\label{sec:approach}
To derive our incident and vulnerability \taxonomy, we moved from definition synthesis and usability methodology to iterative internal and external feedback rounds. 

\textbf{Definitions.} Four authors worked jointly in revising common definitions for security, AI, AI system, etc, to define an AI security incident and thereby the scope of the \taxonomy.

\textbf{Usability methodology.} In parallel, we relied on methodology from usability research: all necessary concepts will emerge from a small sample~\cite{hwang2010number}. We organized 
twelve calls (30-60 minutes) with distinct seasoned subject matter experts (working in AI red teaming, security industry, AI governance, or market oversight), inquiring about their ``wish list'' for a reporting scheme (mid-2024). We considered all emerging concepts and matched them across experts, with model card approaches, 
across existing databases, 
and with AI security literature.

\textbf{Feedback rounds and pilot deployment.} We went through several rounds of internal and external feedback. This included deploying the taxonomy in different contexts, tagging individual incidents up to an entire database.

\textbf{Future work.} Subsequently (late 2025), the proposal was selected as the basis for an ETSI security information exchange standard. During this time, two more experts provided in-depth feedback and became authors. As part of this work, we are (early 2026) collecting feedback from another 20 experts to adapt our \taxonomy to agentic AI systems, which is part of future work. This work also involves coding of agentic incidents as listed in OWASP\footnote{\url{https://genai.owasp.org/resource/owasp-top-10-for-agentic-applications-for-2026/}}. During this process, one more author joined this paper.

\section{Background}\label{sec:background}
We first introduce needed definitions and revisit related approaches describing AI systems, non-AI security incident reporting, and non-security AI incident reporting.

\begin{table*}[h!]
\caption{Existing databases and collections of AI (security) incidents. We list the kind of entity (Government, Industry, Non-Profit, and Academia) and number of entries as of July, 2026. We also report whether incidents cover ethics/non-security, security, AI security, or offensive use of AI. We then review which databases contain reproducible incidents and revisit the sources of the incidents: research papers, media, anonymous reports, or platforms where registered users can submit reports. We finally report whether the databases are machine-readable and link to the CVE databases.}\label{tab:AISEcCollections}
\centering
\resizebox{\textwidth}{!}{
\begin{tabular}{@{}lcllllllllllllllllr@{}}
\toprule
&&&&\multicolumn{4}{c}{Incident Type}&&&&\multicolumn{3}{c}{Sources} \\ 
\cmidrule(l){5-8}\cmidrule(l){12-15}
Database &&\# Entries && \rotvertical{Ethical} & \rotvertical{Security} & \rotvertical{AI Security} & \rotvertical{Offensive}\hspace{0.3em}\rotvertical{AI usage} &&\rotvertical{Reproducible}\hspace{0.3em}\rotvertical{Incidents} && \rotvertical{Research}\hspace{0.3em}\rotvertical{Paper} & \rotvertical{Media} & \rotvertical{Submissions} & \rotvertical{Anonymous}  && \rotvertical{Machine}\hspace{0.3em}\rotvertical{readable} & \rotvertical{Direct}\hspace{0.3em}\rotvertical{Integration} & \rotvertical{Links to} \\  
\midrule
\citet{oecdaim} AIM & G &  $>$16,600 && \checkmark & \checkmark & \checkmark & \checkmark && && \checkmark & \checkmark & & & & CSV\\
\citet{huntr} & I & $>$2,300 && & \checkmark & &&& \checkmark & &&&\checkmark  &  && -- && CVE \\
AIRD~\cite{airisk} &I& $>$2,000 && & & \checkmark  &&& \checkmark&  & &&\checkmark&  & & CSV & Python & CVE \\
AVID~\cite{avid} & NP & $>$1,500  &&  \checkmark & \checkmark & \checkmark & \checkmark &&  && \checkmark & \checkmark & & & & JSON & Python & CVE\\
AIID~\cite{mcgregor2021preventing} & NP & $>$1,500 && \checkmark & \checkmark & \checkmark & \checkmark &&&& \checkmark & \checkmark & & && CSV & MongoDB \\
Deepfake DB~\cite{deepfakedb} & A &  200 && & & & \checkmark && && & \checkmark & & && CSV \\
\citet{lve} & A & 136 && \checkmark & \checkmark & \checkmark & \checkmark && \checkmark && \checkmark & \checkmark &\checkmark&  && JSON & Python & \\
\citet{grosse2023your} & A & 32 && \checkmark & \checkmark & \checkmark &&& & &&  &  &\checkmark&& --\\
\citet{mitre} case studies &G&  31 && \checkmark & \checkmark & \checkmark & \checkmark && && \checkmark & \checkmark & & && JSON & Python & CVE \\
\bottomrule
\end{tabular}%
}
\end{table*}

\subsection{Definitions}\label{sec:defs}
We base our \taxonomy on core terminology from existing standards and use them to construct a definition of ``AI security incident'' and ``vulnerability'', the two core concepts of our taxonomy. 
The complete list of all required definitions is in the Appendix. Here, we only revisit the four most important definitions: an ``AI system'', ``security'', and based on these an ``AI security incident'' (the target of our collection), and lastly a ``vulnerability'' (underlying the incident).
We first start with the definition of the ``AI system'', derived from the \citet{OECDAiSystemDefMemo2024} definition.

\begin{definition}
\label{def:aisys}
\textbf{AI system.} \textit{An AI system is a machine-based system that, for explicit or implicit objectives, infers, from the input it receives, how to generate outputs such as predictions, content, recommendations, or decisions that can influence physical or virtual environments. Different AI systems vary in their levels of autonomy and adaptiveness after deployment.}
\end{definition}

We look to this definition of ``AI system'' as a foundation for our work for two main reasons. Firstly, the relevant OECD recommendation also identifies that ``AI systems'' and the ``AI models'' they embed are distinct. This allows a potential AI security incident report to identify if an exploit is scoped strictly to the ``mathematical construct'' that is a model, or interacts with the larger AI system. Secondly, the definition is used by or influences those of other intergovernmental organizations (such as the United Nations) and prominent market regulations relevant to incident reporting (such as the EU AI Act) \cite{OECDAiSystemDefMemo2024}.

We also recognize ``security'' as broadly defined in ISO/IEC 27000 by~\citet{isosecruityRef}

\begin{definition}\label{def:security}
\textbf{Security.} \textit{The preservation of confidentiality, integrity, availability, accountability, authenticity, and reliability of a system}.
\end{definition}

We further additionally review the definitions of confidentiality, integrity, and availability, also known as the ``CIA triad''. This well established concept is also used in scientific work covering AI security~\cite{biggio2018wild}.

\begin{definition}
\textbf{Confidentiality.} \textit{Property that information is not made available or disclosed to unauthorized individuals, entities, or processes \cite{isosecruityRef}.}
\end{definition}

\begin{definition}
\textbf{Integrity.} \textit{Property of accuracy and completeness \cite{isosecruityRef}.}
\end{definition}

\begin{definition}
\textbf{Availability.} \textit{Property of being accessible and usable on demand by an authorized entity \cite{isosecruityRef}.}
\end{definition}

Importantly, this definition of ``security'' based on the CIA triad is shared by other ISO/IEC standards specific to AI.\footnote{For example: ISO/IEC 42001 Information technology — Artificial intelligence — Management system; ISO/IEC TR 24028 Information technology — Artificial intelligence — Overview of trustworthiness in artificial intelligence; ISO/IEC TR 24368 Information technology — Artificial intelligence — Overview of ethical and societal concerns}
Based on these foundational terms, we define an ``AI security incident'', extending the existing definition of ``AI incident'' by the~\citet{OECDAIinc2024} and considering the above CIA security triad aspects shared across ISO/IEC standards. 

\begin{definition}
\label{def:inj}
\textbf{AI security incident.} \textit{An AI security incident is an event, circumstance, or series of events where the confidentiality, integrity, or availability of an AI system is affected and which leads to any of the following harms: (a) injury or harm to the health of a person or groups of people; (b) disruption of the management and operation of critical infrastructure; (c) violations of human rights or a breach of obligations under the applicable law intended to protect fundamental, labor and intellectual property rights; (d) harm to property, communities or the environment.}
\end{definition}

We use these definitions to avoid, for example, confusion between general security and AI-specific security incidents~\cite{bieringer2022industrial}. Analogous to security, AI security incidents usually exploit a vulnerability present in the system. 

\begin{definition}
\textbf{AI Vulnerability.} \emph{A weakness of an asset or a measure to reduce risk that can be exploited as a potential cause of an unwanted AI security incident, which can result in harm to a system or organization}, \emph{from ISO/IEC 27000:2018~\cite{isosecruityRef}}.
\end{definition}

There is an $n$ to $m$ relationship between vulnerabilities and incidents: Multiple incidents can arise from a single vulnerability, and a single incident can be caused by one or more vulnerabilities. Our taxonomy allows to denote AI vulnerabilities and AI security incidents separately, yet in relation to each other, as depicted in Figure~\ref{fig:template}. 

Many AI vulnerabilities in the form of AI security research exist~\cite{biggio2018wild, cina_wild_2022,gu2019badnets,lei2023new,oliynyk2023know,shen2023anything,shumailov2021manipulating,yao2024survey}. Established frameworks, like the MITRE ATLAS\footnote{\url{https://atlas.mitre.org/matrices/ATLAS}}, or research frameworks~\cite{biggio2018wild} systematize these vulnerabilities. Information such as required access patterns~\cite{biggio2018wild,oliynyk2023know,huang2011adversarial} is crucial for systematization. Below, we provide corresponding references whenever such information is in the taxonomy. 

\subsection{Describing AI systems}\label{sec:modelCards}
To describe an AI security incident accurately, the impacted AI system must be reported. 
\citet{mitchell2019model} provided an initial proposal for such a description. Afterwards, model providers and various stakeholder groups used these approaches to access information about a model and conclude corresponding action. Derivative templates are fact sheets (\citet{arnold2019factsheets}), datasheets for datasets (\citet{gebru2021datasheets}), method cards (\citet{adkins2022method}), risk cards (\citet{derczynski2023assessing}), use case cards (\citet{hupont2024use}), and AI cards (\citet{golpayegani2024ai}).

However, while most of these approaches require the documentation of general metadata about the model (e.g., model type or model version), their content related to security is often limited and only covers implemented security measures, if at all. Yet, they provide suitable guidance on how to describe the underlying system affected by the incident. Where reasonable, we thus link the proposed elements of our \taxonomy  (Section~\ref{sec:TaxSecTwo}) with this body of literature.

\subsection{AI security incident reporting frameworks}\label{sec:AIincidents}
Our work benefits from other collections of AI-related incidents, and summarize such efforts in Table~\ref{tab:AISEcCollections}.
Although there are reports of real-world AI security incidents as discussed in the introduction~\cite{grosse2023your,mcgregor2021preventing,avid}, there are few incident databases cover AI security. As is further visible, most databases do not rely on first-party submitted reports but on publicly available third-party media articles. This will change when reporting for AI becomes mandatory, for example under the EU AI Act. Existing databases cover more than AI security, and include ethical incidents related to AI. 
An incident involving unethical application of AI is, however, fundamentally different from an AI security incident due to the absence of an attacker. 
Still, many databases provide guidelines on collected information and emphasize other relevant features. These include machine readability (seven of nine databases allow direct export of data), links to other vulnerability databases like CVE (four of nine), and lastly direct integration via, for example, APIs with programming languages for data analysis (five of nine). The latter is especially relevant to research, where corresponding frameworks in non-AI security have enabled more detailed analysis of the threat surface~\cite{neuhaus2010security,li2017mining}.

To benefit from the knowledge within these databases, we use them to motivate design choices in our \taxonomy below.

\subsection{Security incident reporting frameworks}\label{sec:secReporting}
Another area to draw knowledge from is cybersecurity incident reporting. 
Frameworks for collecting, classifying, and reporting cybersecurity incidents began to take shape in 1999 with the seminal work on the Common Vulnerability and Exposures (CVE) database~\cite{baker1999cve}. The CVE Program~\cite{cveprogram} has since matured and achieved widespread adoption. Building on the success of CVE, complementary initiatives evolved, for example, assessing the severity of a vulnerability~\cite{nistcvss,scarfone2009cvss}, and collections of attack tactics and techniques~\cite{mitreattack,mitreattackcve}. 
While CVEs may still affect AI---CVE-2023-25661,\footnote{\url{https://nvd.nist.gov/vuln/detail/CVE-2023-25661}}, security vulnerabilities and AI security vulnerabilities are fundamentally different~\cite{ai2023artificialAppB,cattell2024coordinated,bieringer2024position}.
Our AI security incident \taxonomy is specifically designed to address security incidents related to AI models.

Our approach mirrors the CVE model, and there are lessons to be learned for AI security reporting. For instance, as the volume of AI security incidents grows steadily, a labeling schema is required to uniquely identify incidents that are classified in our proposed taxonomy. Additionally, the process of reviewing and labeling incidents could become resource-intensive. Therefore, a governance body — similar to the CVE Numbering Authorities (CNA)~\cite{cna} — must be established to oversee this task.

If our taxonomy gains acceptance as an ETSI standard, further initiatives can be explored. This includes, but is not limited to, integrating the CVE and our taxonomy, assigning severity scores similar to CVSS, or establishing a procedure to publish vulnerabilities and incidents.  

\section{AI security incident \taxonomy}\label{sec:dbtaxonomy}
Our \taxonomy consists of a three-part description of vulnerabilities (V1-V3) and a four-part description of incidents (I1-I4). Both share the metadata (V1/I1) and the appendix (V3/I4). A vulnerability encompasses a description of the AI system and the vulnerability (V2), whereas the incident entails a description of the incident (I2) and its implications (I3). We now revisit the individual questions and references in each part. An overview of the taxonomy is in 
Figure~\ref{fig:template}, the full version with all questions and definitions is in the appendix in Table~\ref{tab:Taxpart1}-\ref{tab:Taxpart3}.

\begin{figure}
\centering
\vspace{-0.2em}
\includegraphics[width=\linewidth,scale=1.5]{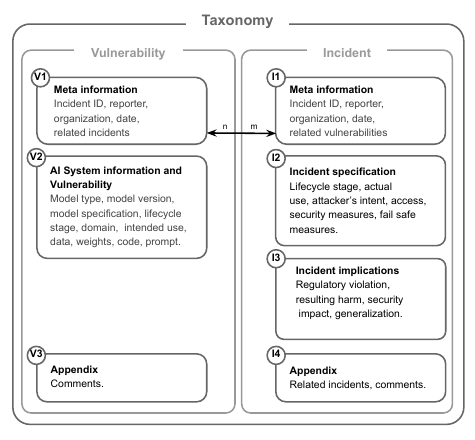}
\vspace{-1.5em}
\caption{Our proposed taxonomy for reporting AI security incidents. Details of the incident and the implicated vulnerabilities in the AI system are separated, given that vulnerabilities may exist independent of the one or many incidents to which they may contribute.}\label{fig:template}
\vspace{-0.2em}
\end{figure}

\subsection{Metadata (\refTax{1}{1}-\refTax{1}{5})}\label{sec:MetaData}
The metadata is a short block containing the vulnerability's or incident's high-level information, and is, under variations, part of many existing databases~\cite{huntr, mcgregor2021preventing,lve,avid,oecdaim}. This information forms the first section with \taxonomy elements \refTax{1}{1}-\refTax{1}{5}.

We opted to assign an incident ID, a name or ID of the incident reporter, their organization, and the alleged date (or interval, or unknown) of the incident. It is important to note that vulnerabilities or incidents may be reported anonymously. Although, as stated above, any question of our proposal can be dropped, we want to emphasize that the reporter and their organization may be pseudonymized or kept fully anonymous, depending on the context of usage. Vulnerabilities (\refVuln{1}{5}) and incidents (\refInc{1}{5}) are linked to each other, where each can be linked to one or more of the other. This $n$ to $m$ relationship is also depicted in Figure~\ref{fig:template}.

In general, it may be desirable that a vulnerability is specified for a given incident: Several models may encompass a vulnerability, yet only one was exploited. It may thus be desirable to be able to overwrite the information of a vulnerability given a specific incident. 

\subsection{Targeted AI system and Vulnerability (V2)}\label{sec:TaxSecTwo}
The vulnerability is strongly related to an AI system, which we describe in detail. More specifically, we document information about the model, its intended use and supply chain, application-specifics, and finally the vulnerability itself as enabled attack and target asset. Ideally, much of this information is imported from a model card or similar documentation to reduce documentation work.

\textbf{Model information (\refVuln{2}{1}-\refVuln{2}{3}).} We collect basic information about the model. This includes the model type, whether a model is generative or predictive; a specific reply here could be a support vector machine. Such information was suggested by half of our experts and is used in several model card approaches~\cite{mitchell2019model,arnold2019factsheets,adkins2022method}. We then collect more specific information like the exact version of the model, information also encompassed in many model cards~\cite{mitchell2019model,adkins2022method} and requested by a fourth of our experts. Lastly, if possible, the model should be specified as detailed as possible (e.g., model weights), to enhance reproducibility, as suggested by another quarter of our experts. 

\textbf{Intended use and model supply chain (\refVuln{2}{4}-\refVuln{2}{7}).}
We document the intended use, loosely following the EU AI Act~\cite{madiega2021artificial}, and as in many model card approaches~\cite{mitchell2019model,hupont2024use,arnold2019factsheets,adkins2022method}.
Furthermore, it is crucial to understand possible (external) influences on the AI system stemming from the supply chain, as also suggested by two of our experts and in several AI model card approaches~\cite{mitchell2019model,gebru2021datasheets,adkins2022method}. Following scientific literature in AI security, we opted to divide the supply chain into three parts: data~\cite{mir2024transparency},  model weights~\cite{gu2019badnets}, and code~\cite{grosse2019adversarial, shumailov2021manipulating}. Any of these may allow the import of security vulnerabilities. Especially for data, prior work suggests which information could be collected~\cite{benderDriedman2018Data}. 

\textbf{Hardware and Application (\refVuln{2}{8}-\refVuln{2}{9}).}
In addition, we propose to track the hardware, as suggested by one expert. Moreover, AI vulnerabilities may need specific details to be reproducible or understandable. An example is collecting the system prompt when using a large language model (LLM). Such information was suggested by one expert and is collected in LLM-related databases~\cite{lve}, and model cards~\cite{derczynski2023assessing}. 

\textbf{Vulnerability Information (\refVuln{2}{8}-\refVuln{2}{9}).}
Finally, we recommend recording the attack. For our tagging, we relied on the categorization by \citet{grosse2023your}. Any other framework, however, may also serve this purpose. An AI security incident is further categorized by the resources used by the attacker to carry out the attack~\cite{apruzzese2023real}. We thus recommend identifying and reporting the target asset, e.g., whether the model or the entire system was targeted, as pointed out in prior work~\cite{bieringer2022industrial}.

\subsection{Incident information (I2)}\label{sec:TaxSecThree}
Orthogonally to the vulnerability, we describe the incident. To this end, we collect information about the AI system at the time of the incident, relevant involved components, defensive measures, and the attacker's intent. 

\textbf{Domain, lifecycle stage, and actual use (\refInc{2}{1}-\refInc{2}{3}).}
Understanding the context of an AI security incident is crucial. 
As suggested by the CSET AI Harm Framework~\cite{csetFramework} annotation guide, we collect the AI system's deployment domain~\cite{csetAnnotation}. Intuitively, a deployment in physical domains has different implications than in the digital space. 
We also collect lifecycle information as specified by ISO\footnote{ISO/IEC 22989 Information technology — Artificial intelligence — Artificial intelligence concepts and terminology, \url{https://www.iso.org/standard/74296.html}}, as suggested by two of our experts. Also in research, there are calls to include lifecycle information into security analysis, such as red-teaming~\cite{majumdar2025red}. Finally, and if diverging from the intended use~\cite{madiega2021artificial}, we advocate denoting a usage during the incident.

\textbf{Relevant components and defensive measures (\refInc{2}{4}-\refInc{2}{6}).}
Access rights to all components of the AI system (data, model, queries, etc) should be reported to identify how the attacker possibly entered the system. The latter was suggested by one expert and studied in prior work~\cite{grosse2023Towards,grosse2023machine}.
Access patterns also give an understanding which access may have prevented the attack. Along these lines, there is value in understanding what measures were in place before the incident took place, but did not prevent it. To this end, we advocate reporting existing security measures, as suggested by a quarter of our experts and the literature in model cards~\cite{hupont2024use,arnold2019factsheets,adkins2022method}. This is in particular relevant for large language models, where implemented guardrails should be reported~\cite{rebedea2023nemo}. Along the same lines, we suggest also collecting fail-safe measures.

\textbf{Attacker's Intent (\refInc{2}{7})}
To ensure that an incident is indeed an AI security incident as defined in Definition~\ref{def:inj} and Table~\ref{tab:Taxpart2}, the attacker's malicious intent should be recorded. This was suggested by half of our experts, forms part of several databases~\cite{avid,oecdaim,deepfakedb}, and was suggested by~\cite{grosse2023your}. 

\subsection{Incident implications (I3)}\label{sec:TaxSecFour}
To understand the severity of an AI security incident, it is crucial to monitor resulting damage. We capture such effects in this brief section with elements \refInc{3}{1}-\refInc{3}{4}. The information in this section may be difficult to derive depending on the role of the reporter---a possible solution for this problem is to rely on a committee or internal risk management teams. 

As the first element, we propose to monitor whether the incident led to non-compliance with existing regulations, a factor brought up by industry participants in security contexts~\cite{bieringer2022industrial,grosse2023machine}. Secondly, any harm that occurred should be well-documented, as suggested by the CSET AI Harm Taxonomy~\cite{csetFramework}  applied within the AI Incident Database~\cite{mcgregor2021preventing}. While 
determining (or defining) the extent of harm of an AI incident can be inherently uncertain~\cite{lessonsAIIncidents}, we advocate that it is still worth attempting to record harms resulting from security incidents. 

A different level of abstraction is to use the CIA definition 
to characterize the impact of the incident. Finally, and beyond the immediate incident, the goal of our \taxonomy is to understand vulnerabilities beyond an individual system. We thus advocate explicitly reporting an assessment of whether the incident could affect other AI systems. This was also suggested a fourth of our experts and is implemented in the Huntr database already~\cite{huntr}.

\section{Pilot Deployments}
The taxonomy, especially as a standard, has to be usable. We incorporated this into the design process, going through several pilot deployments while developing. 

First discussions rounds among four of the authors led to duplicate removal. For example, some experts raised the question of model reuse, while others noted the relevance of keeping track of supply-chain-related information regarding the model.
In these early discussions, meta-information similar to the CSET AI Harm Taxonomy~\cite{csetAnnotation} like incident date, annotator, and whether the AI system was a cyber-physical system was added.  

We then applied the taxonomy to incident descriptions.  
Two of the authors did an initial coding round of 16 incidents.
This led to several changes in the questions, sections, and wording of question explanations. Afterwards,
three of the authors used the resulting, improved \taxonomy to tag five incidents from existing databases. 
Due to the open nature of the tags, we compute the Jaccard index per incident across coders. The index ranges between 0.65 and 0.78 (mean 0.71), indicating good agreement and that the coders capture similar topics. 
The largest disagreement was in fields regarding harm classification, where two authors had significantly less background than the third, showing the need for guidelines. 
Other disagreeing examples related to the generalizability of incidents, a challenging question for which, to the best of our knowledge, standard assessment procedures are not yet established.

Encouraged by these results, we shared the proposal, as an early version of this paper, with the initial and additional experts for a broader, final round of feedback (early 2025). While several experts promised to apply the taxonomy to internal incidents, no significant changes to the taxonomy emerged. During the writing of this paper, four of the experts became co-authors. 
Lastly, we investigated whether an LLM (ChatGPT) can be used for incident filing. Using the AIID~\cite{mcgregor2021preventing}, we observed promising results. However, the agreement with the human coders is significantly lower, around 0.55. LLMs should thus only be used to support human annotation; other forms of automation like automated filling based on model cards should be relied on. 

\section{Limitations}
We synthesized the above \taxonomy from a diverse set of practitioners to develop a solution that addresses real-world needs and challenges.  
Early feedback indicated that the \taxonomy is bureaucratic, asking for information not known to the reporter, or not desirable to exchange outside the company. We thus highlight again that a subset of the \taxonomy may be desirable over the complete approach in some cases. However, other properties appeared on the community's wishlist, including usability, machine readability, and a proper treatment of the AI model as intellectual property of the owner, which we discuss more in depth in the appendix.

Moreover, our taxonomy does not cover offensive use of AI, AI agents, multi-agent systems, AI systems with multiple AI models or components, or recursive usage of AI. While we are working on such an extension, it is beyond the scope of this work.
In general, the \taxonomy and underlying definitions will have to be adapted for new use cases and trends in AI. Furthermore, AI and its security incidents are very application-specific\footnote {As an example, consider AlphaFold's prohibited use policy \url{https://github.com/google-deepmind/alphafold3/blob/main/WEIGHTS_PROHIBITED_USE_POLICY.md}}, and it may be necessary to adapt the \taxonomy for specific applications. Along these lines, determining which fields are optional or mandatory is still an open research question. In addition, determining whether an incidents generalizes from one model to another is hard, and currently an open research questions beyond the insight that most models are vulnerable to most attacks. Especially challenging is that an attack may rely on multiple steps or inputs, and that it is also currently unknown how to monitor AI systems to spot the largest fraction of attacks based on constrained resources.

\section{Conclusion and Future Work}
We presented the derivation of an AI security incident collection taxonomy based on practitioner input and related scientific work. Beyond the split between vulnerability and incident, we suggest concrete information to be collected. To ensure the taxonomy is applicable, we ran several promising pilot deployments that show high inter-coder agreement across users. We evaluated whether modern LLMs can operationalize a harm taxonomy at scale, but found limitations regarding coding agreement.

To enable or increase practical usage of the taxonomy, several open research questions have to be tackled. This includes the identification of the attacker. While similar frameworks such as CVE neither cover the identification nor the collection of information on the attacker, in the long run, approaches to obtaining such information would greatly benefit both research, AI deployers, and stakeholders. We have also outlined that in some cases, like supply chain (\refVuln{2}{5}-\refTax{2}{7}), attacks (\refVuln{2}{10}), and harm (\refInc{3}{2}),  a decision has to be taken on which frameworks are to be chosen. This is an additional avenue for future work. Finally, usability studies could help to understand and advance both the usefulness of the collected reports~\cite{winckler2013identifying,grant2015save}. 

\section*{Acknowledgments.}
We would like to thank our anonymous experts and their investment in giving initial inputs and feedback on this work. Some of our experts consented to being explicitly named, including
 Hyrum Anderson, Cisco; Leon Derczynski, NVIDIA Corporation, whom we thank for his feedback on our definitions, methodology, and references; and Yuvaraj Govindarajulu, AIShield, Bosch Global Software Technologies; 
 and Brian Pendleton, whom we thank especially for his feedback regarding CVE/CWE. 

\bibliography{lit}

\appendix

We base our \taxonomy on core terminology from existing standards and use them to construct a definition of ``AI security incident'' and ``vulnerability'', the two core concepts of our taxonomy. 
The complete list of all required definitions is in the Appendix. Here, we only revisit the four most important definitions: an ``AI system'', ``security'', and based on these an ``AI security incident'' (the target of our collection), and lastly a ``vulnerability'' (underlying the incident).
We first start with the definition of the ``AI system'', derived from the \citet{OECDAiSystemDefMemo2024} definition.

\begin{definition}
\label{def:aisys}
\textbf{AI system.} \textit{An AI system is a machine-based system that, for explicit or implicit objectives, infers, from the input it receives, how to generate outputs such as predictions, content, recommendations, or decisions that can influence physical or virtual environments. Different AI systems vary in their levels of autonomy and adaptiveness after deployment.}
\end{definition}

We look to this definition of ``AI system'' as a foundation for our work for two main reasons. Firstly, the relevant OECD recommendation also identifies that ``AI systems'' and the ``AI models'' they embed are distinct. This allows a potential AI security incident report to identify if an exploit is scoped strictly to the ``mathematical construct'' that is a model, or interacts with the larger AI system. Secondly, the definition is used by or influences those of other intergovernmental organizations (such as the United Nations) and prominent market regulations relevant to incident reporting (such as the EU AI Act) \cite{OECDAiSystemDefMemo2024}.

We also recognize ``security'' as broadly defined in ISO/IEC 27000 by~\citet{isosecruityRef}

\begin{definition}\label{def:security}
\textbf{Security.} \textit{The preservation of confidentiality, integrity, availability, accountability, authenticity, and reliability of a system}.
\end{definition}

We further additionally review the definitions of confidentiality, integrity, and availability, also known as the ``CIA triad''. This well established concept is also used in scientific work covering AI security~\cite{biggio2018wild}.

\begin{definition}
\textbf{Confidentiality.} \textit{Property that information is not made available or disclosed to unauthorized individuals, entities, or processes \cite{isosecruityRef}.}
\end{definition}

\begin{definition}
\textbf{Integrity.} \textit{Property of accuracy and completeness \cite{isosecruityRef}.}
\end{definition}

\begin{definition}
\textbf{Availability.} \textit{Property of being accessible and usable on demand by an authorized entity \cite{isosecruityRef}.}
\end{definition}

Importantly, this definition of ``security'' based on the CIA triad is shared by other ISO/IEC standards specific to AI.\footnote{For example: ISO/IEC 42001 Information technology — Artificial intelligence — Management system; ISO/IEC TR 24028 Information technology — Artificial intelligence — Overview of trustworthiness in artificial intelligence; ISO/IEC TR 24368 Information technology — Artificial intelligence — Overview of ethical and societal concerns}
Based on these foundational terms, we define an ``AI security incident'', extending the existing definition of ``AI incident'' by the~\citet{OECDAIinc2024} and considering the above CIA security triad aspects shared across ISO/IEC standards. 

\begin{definition}
\label{def:inj}
\textbf{AI security incident.} \textit{An AI security incident is an event, circumstance, or series of events where the confidentiality, integrity, or availability of an AI system is affected and which leads to any of the following harms: (a) injury or harm to the health of a person or groups of people; (b) disruption of the management and operation of critical infrastructure; (c) violations of human rights or a breach of obligations under the applicable law intended to protect fundamental, labor and intellectual property rights; (d) harm to property, communities or the environment.}
\end{definition}

We use these definitions to avoid, for example, confusion between general security and AI-specific security incidents~\cite{bieringer2022industrial}. Analogous to security, AI security incidents usually exploit a vulnerability present in the system. 

\begin{definition}
\textbf{AI Vulnerability.} \emph{A weakness of an asset or a measure to reduce risk that can be exploited as a potential cause of an unwanted AI security incident, which can result in harm to a system or organization}, \emph{from ISO/IEC 27000:2018~\cite{isosecruityRef}}.
\end{definition}

There is an $n$ to $m$ relationship between vulnerabilities and incidents: Multiple incidents can arise from a single vulnerability, and a single incident can be caused by one or more vulnerabilities. Our taxonomy allows to denote AI vulnerabilities and AI security incidents separately, yet in relation to each other, as depicted in Figure~\ref{fig:template}. 
\begin{table*}[]
\caption{Reporting scheme with description for annotators and examples. Part shared between vulnerability (V$n.n$) and Incidents (I$n.n$): Meta information and Appendix.}\label{tab:Taxpart1}
\footnotesize
\begin{tabular}{p{.2cm} p{1.2cm} p{2.7cm}p{2.3cm}p{6cm}p{3cm}}
\toprule
 &  & \textbf{Short name} &  \textbf{Long name or question}              & \textbf{Description for annotators} & \textbf{Example}  \\
 \midrule
 \multirow{4}{*}{\rotvertical{Meta Information}} & \refTax{1}{1}   & Incident ID     & Incident ID     & The unique identifier of the incident. & ID\_V\_1716 / ID\_I\_1716         \\
  & \refTax{1}{2}   &  Incident Reporter         &  Name of incident reporter      &  Name or also username from within recording.                              & John Doe / AIIncidentreporter001     \\
  & \refTax{1}{3}   & Reporting Organization                            & Name of reporting organization                         &  The legal name of the reporting organization.                             & Acme Inc.        \\
  & \refTax{1}{4}   &  Incident Date             &  Date the incident occurred      &  Date or Time interval, if unkown estimate interval                          &  13/12/2024   \\
  & \refVuln{1}{5} & Caused Incidents & Incidents caused by Vulnerability & A List of incidents this vulnerability is associated with & ID\_I\_1234 \\
  & \refInc{1}{5} & Associated Vulnerabilities & Vulnerabilities that likely caused the incident & One or more vulnerabilities that caused the incident & ID\_V\_4321 \\
 \addlinespace 
\multirow{6}{*}{\rotvertical{Appendix}}                               
& \refInc{4}{1}   & Links to other incidents   & Links to similar / same cases in other /same database(s)    & In case you are aware of a incidents that has been reported in another database or that is similar to this incident in this database & This incident has had similar consequences like AIID incidents 38 or simply 'URL'   \\  
& \refVuln{3}{1}/\refInc{4}{2}   & Comments   & Additional comments.                                   & Anything you'd like to add which has not been covered in the previous fields.               & The attack does not work on model ABC1.2 and ABC1.4, but only on ABC1.3   \\  

 \bottomrule
  \end{tabular}
\end{table*}

  \begin{table*}[]
\caption{Reporting scheme with description for annotators and examples. Part depicted describes uniquely vulnerabilities (\refVuln{2}{1}-\refVuln{2}{11}). \refVuln{2}{9} denotes an example for a technique-specific field, here a large language model. }\label{tab:Taxpart2}
\footnotesize
\begin{tabular}{p{.2cm} p{.4cm} p{2.7cm}p{2.5cm}p{6cm}p{3cm}}
\toprule
 &  & \textbf{Short name} &  \textbf{Long name or question}              & \textbf{Description for annotators} & \textbf{Example}  \\
 \midrule
  \multirow{45}{*}{\rotvertical{AI System Information and Vulnerability}} & \refVuln{2}{1}   & Model type      & Specify the ML model type.                             &  Prediction, generative, but also more concrete, Transformer, DNN, ConvNet, SVM etc                      &  Support Vector Machine                              \\
  & \refVuln{2}{2}   & Model version   & High level identifier, as noted in model card.         &  This identifier should allow for correct identification of the specific model version.        &  Llama3.2:1B  \\
  & \refVuln{2}{3}   & Model specification                               & Provide as precise information on the model as possible.                   & Information could contain URLs to or files with weights, full model files or model cards.         & 'URL'                          \\
  & \refVuln{2}{4}   & Intended use    & Described the intended use of the AI system.           & Intended purpose means the use for which an AI system is intended by the provider, including the specific context and conditions of use, as specified in the information supplied by the provider in the instructions for use, promotional or sales materials and statements, as well as in the technical documentation.                      & Object detector to tag cats on image, Chatbot to reply to questions about air plane fares, etc  \\
  & \refVuln{2}{5}   & Model supply chain - Data                         & Was publicly available data used? If yes, from where, was data extended or adapted and if yes how?                & Specify dataset with name and URL, as well as any modifications that the data underwent before being used.            & For example: Model used MNIST data from 'URL', where we added 1000 images that were handlabelled by our team.                          \\
  & \refVuln{2}{6}   & Model Supply Chain - Model weights                & Are the model weights derived from an existing model, and if yes, which one and how were they adapted?            & If the original design or weights have a name, provide this name or possibly a URL. Further describe all steps the model underwent to adapt it to the current status.          & For example: Based on Llama3.2:1B, but finetuned on the above described dataset for n iterations using optimizer o and loss l.         \\
  & \refVuln{2}{7}   & Model Supply Chain - Code                         & What public code was used to generate and run the model?                   & Provide libraries used with their version name, and also enumerate github repositories with link and access data or other re-used code from the project.   & For example: Pytorch version 2.4.1. and Code from github repo 'this url' was used, forked January 1st 2024.                            \\
    & \refVuln{3}{8}   & Hardware   & What hardware was used, and could it have affected the incident?           & Please describe the specific hardware used on the device or server the application ran, and whether the setting implies that the hardware played a role in the incident.         & Mac M1, could not have affected the incident OR could have affected the incident via a glitch.  \\
    & \refVuln{2}{9}  & System prompt\textbf{*}   & If the system is a generative AI system, provide the system prompt.        & If known, report the entire system, prompt as used in the setting where the incident occurred.    & For example: 'please be polite to the user and don't hallucinate. The user writes: '            \\
   & \refVuln{2}{10}   & Attack    & Specify the concrete type of attack that has been carried out.             & Any attack method, technique, framework, or tool. 
   & Poisoning        \\
  & \refVuln{2}{11}  & Model exclusive Vulnerability  & Did the attack target the model explicitly and the model only?        & Based on the definition in Sect.~\ref{sec:defs}                 & Yes, model was targeted.         \\
 \bottomrule                        
\end{tabular}
\end{table*}

\begin{table*}[]
\caption{Reporting scheme with description for annotators and examples, Incident description (\refInc{2}{1} - \refInc{3}{4}).}\label{tab:Taxpart3}
\footnotesize
\begin{tabular}{p{.2cm} p{.4cm} p{2.3cm}p{2.7cm}p{4.5cm}p{4.5cm}}
\toprule
 &  & \textbf{Short name} &  \textbf{Long name or question}              & \textbf{Description for annotators} & \textbf{Example}  \\
 \midrule 
 \multirow{35}{*}{\rotvertical{Incident specification}} &  \refInc{2}{1}   &  Physical Domain                    & Did the incident occur in a physical environment?     & Did the incident affect an AI that is interacting with physical objects or otherwise physically with humans?          & Yes, pedestrian not detected.        \\
 & \refInc{2}{2}   & Lifecycle stage                                   & Define the stage of the AI lifecycle which the model was in when the incident occurred.                            & Stages of the AI lifecycle are: Inception, Design \& Development, Verification \& Validation, Deployment, Operation and Monitoring, Re-evaluate, Retirement.  & Deployment       \\
 & \refInc{2}{3}   & Actual use during incident, if differing from 2.6 & Describe the use of the AI system during the incident, if differing from intended use (\refVuln{2}{x})                      & This may mean the usage for which an AI system was not intended by the provider, including the specific context and conditions of the actual use.            & Used to tag foxes instead of cats, used to get legal advice model was not trained for.          \\
  & \refInc{2}{4}   & Access     & Which access was given to whom for model, inputs, training, test or evaluation data?                              & Specify which levels of access were given to which users regarding each component.              & Training and evaluation data as well as model was only available to internal model team. The model could be queried via API by the client, but access control prevented any other access by 3rd parties to the model.  \\
  & \refInc{2}{5}   & Security measures                                 & What security measures where used? Where they circumvented and if yes how?      & Give a list of specific security measures that were used including their configurations.          & Adversarial training using FGSM with epsilon 0.1, 0.2 and 0.3 was used on randomly selected 10\% of the dataset were used. Defense was not circumvented as attack was based on L2 norm, not $L_\infty$ norm.            \\
 & \refInc{2}{6}   & Fail safe measures                                & Where there any fail-safe measures implemented?        & Describe any fail-safe measures, such as rebooting the system, not outputting a classification, etc.                    & System fell back on human intervention to prevent further harm.           \\
  & \refInc{2}{7}   & Attacker's Intent                                  & Is there evidence that an entity or a person purposely caused the incident by targeting the AI systems security? & Is the incident an example of an attempt to destroy, expose, alter, disable, steal or gain unauthorized access to or make unauthorized use of the AI system? & 'Yes, the attacker altered the sample until it passed the AI security check' or 'No, the user did not know that the system was not trained on this specific subset of the training data (Disease B). Hence, the system provided a wrong output.' \\
 \addlinespace               \multirow{20}{*}{\rotvertical{Incident Implications}}
  & \refInc{3}{1}   & Regulatory violation                                 & List specific regulatory requirements that are violated due to the incident.    & Please specify any concrete legislation, standard or policy that has been violated by the incidents' occurrence.       & European Union AI Act  \\
  & \refInc{3}{2}   & Resulting harm  & What harm did the incident cause to any participating stakeholder?         & Did the incident lead to any of the following types of harm? Affecting physical health, inducing financial loss, damaging a physical object, affecting intangible property, affecting infrastructure, affecting natural environment, violation of human rights, civil liberties, civil rights or democratic norms.                           & Harm to physical property of all users, mainly food but also some devices due to malfunction of the fridge AI.                         \\
  & \refInc{3}{3}   & Security Impact                                   & Describe how the incident affected the confidentiality, integrity, availability of the AI system.                 & Describe whether the incident caused any leakage of confidential data, lead the system to output a wrong result or to exhibit a bias (integrity), or lead to the system being down or otherwise unusable (availability).                   & During the incident, samples were manipulated to be classified as vehicles and thus harmed integrity.                                  \\
 & \refInc{3}{4}   & Generalization                                  & Does the incident potentially affect other AI systems? & Incident descriptions should allow to determine for a third model if a vulnerability in that model is likely.          & Other AI systems might also be affected if relying on the same dataset "xyz".                   \\                     
\bottomrule                        
\end{tabular}
\end{table*}

\end{document}